# Frequency Following Imaging of Electric Fields from Resonant Superconducting Devices using a Scanning Near-Field Microwave Microscope

Ashfaq S. Thanawalla, B. J. Feenstra, Wensheng Hu, D. E. Steinhauer, S. K. Dutta, Steven M. Anlage, and F. C. Wellstood
Center for Superconductivity Research, Physics Department, University of Maryland, College Park, Maryland 20742-4111.

Robert B. Hammond
Superconductor Technology Incorporated, 460 Ward Drive, Suite F, Santa Barbara, CA 93111-2310.

*Abstract*—We have developed a scanning near-field microwave microscope that operates at cryogenic temperatures. Our system uses an open-ended coaxial probe with a 200 μm inner conductor diameter and operates from 77 to 300 K in the 0.01-20 GHz frequency range. In this paper, we present microwave images of the electric field distribution above a $Tl_2Ba_2CaCu_2O_8$ microstrip resonator at 77 K, measured at several heights. In addition, we describe the use of a frequency-following circuit to study the influence of the probe on the resonant frequency of the device.

## I. Introduction

Low-loss high temperature superconducting microwave filters have significant potential for use in advanced communication systems. Unfortunately, these filters tend to suffer from non-linearity, intermodulation and power-dependent characteristics [1]-[3]. Many of the possible sources of these problems are of local origin. This suggests that local imaging techniques, such as measuring microwave electric fields in operating devices, would be useful. Some of the techniques previously employed include using modulated perturbations [4], transmission line probes [5], scanning force potentiometry [6] or near-field microwave microscopy [7]. In general, however, extracting quantitative local information about resonant structures is difficult to achieve without perturbing the system. In this paper we present images of the local electric field around a superconducting microstrip filter, obtained using a scanning near-field microwave microscope. From these images we determine the perturbation of the open-ended coaxial probe on the resonant frequency of the device being imaged.

## II. Experimental setup

One of the main advantages of using a coaxial probe is the fact that the outer conductor screens components of the electric field that are not normal to the face of the probe. In our microscope, we orient the probe so that we are mainly sensitive to the electric field component normal to the device, $E_z$.

On the other hand, the presence of the probe causes the electrodynamical surroundings of the resonant device under test to change, which consequently tends to modify its resonant properties. We use two different techniques to acquire

Manuscript received September 14, 1998
This work has been supported by the National Science Foundation NSF grant No. ECS-9632811, by the Maryland Center for Superconductivity Research, and by STI.

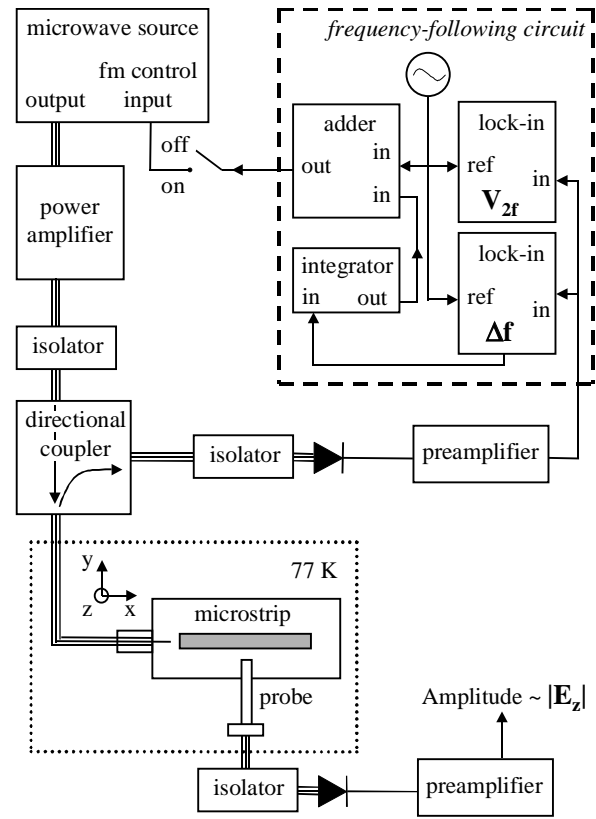

Fig. 1. Schematic of cryogenic microwave microscope. The dotted box indicates the cryogenic section.

our electric field images, a fixed frequency mode and a frequency following mode. In contrast to the fixed-frequency mode, we are able to track the resonant frequency of the device when operating the microscope in the frequency-following mode [8]. This allows us to study the perturbation caused by the probe and to correct our images accordingly.

### A. Fixed-Frequency Mode

Fig. 1 shows our experimental arrangement. An HP 83620B signal generator, which has a frequency range from 0.01 to 20.00 GHz, supplies the microwave power. This power is amplified by an HP 8349B amplifier to about 100 mW and fed to the superconducting microstrip filter through an isolator and a launching pin. The electric field above the microstrip is picked up by the open-ended coaxial probe. Since we operate our microscope in the near-field limit, the spatial resolution is determined by the diameter of the inner conductor of the coaxial probe or the height of the probe above the device, whichever is larger.

As shown in the bottom part of Fig. 1, we measure the electric field distribution by connecting the probe through an isolator to a diode detector, which produces a dc-output proportional to the incident rf-power. The diode output is amplified, filtered, and recorded by a computer, which also controls the sample position using a two-axis cryogenic translation stage. It is important to note that in this configuration we do not connect the frequency following circuit, so that the electric field image is obtained at a fixed source frequency.

### B. Frequency Following Mode

In the frequency-following mode we monitor the signal that is reflected from the input port of the device. The reflected power is extracted using a directional coupler, amplified, and fed to two lock-in amplifiers (see Fig. 1). We use the lock-ins to measure the resonant frequency $f_0$ and the quality factor Q. To do this the frequency of the microwave source is modulated at a rate $f_{FM} = 3$ kHz. One of the lock-in amplifiers is referenced at $f_{FM}$, yielding an output signal that is proportional to the shift of the resonant frequency $\Delta f_0$. This signal is used to keep the source frequency locked to the resonant frequency of the device. The second lock-in amplifier is referenced at $2f_{FM}$, yielding a signal $V_{2f}$ that is proportional to the curvature of the resonance around its peak, and therefore closely related to the quality factor Q. In addition, as in the fixed-frequency mode, we can measure the amplitude of the electric field by employing the signal detected by the probe.

### C. Cryogenic Part

The cryogenic part of the setup, indicated by the dotted box in Fig. 1, is enclosed in a vacuum can which is inserted in a liquid-nitrogen cooled Cryofab Model CSM 6042 Cryostat. Fig. 2 shows the physical arrangement of the cold end. The probe is clamped in an Al-holder that provides the ability to change the probe-sample separation at low temperatures. Behind this plate is situated the Cu-package containing the microstrip resonator. Coaxial cables are connected to both ends of the package to allow us to measure the direct transmission coefficient, $S_{21}$. The xy-motion is established by a slider at the back of the insert, which is connected to vacuum feedthroughs situated on the top of the cryostat.

### III. SAMPLE DESCRIPTION

The images presented in this paper were acquired on a $Tl_2Ba_2CaCu_2O_8$ (TBCCO) superconducting single-pole microstrip resonator. The resonator had a 650 nm thick film of TBCCO deposited on both sides of a 420 µm thick, 4.7 mm X 9.9 mm MgO substrate. One side served as a ground plane, while the other side was patterned into a microstrip with a width of 150 µm and a length L of 7.1 mm. We mounted the resonator in a Cu package and capacitively coupled rf-power to one end.

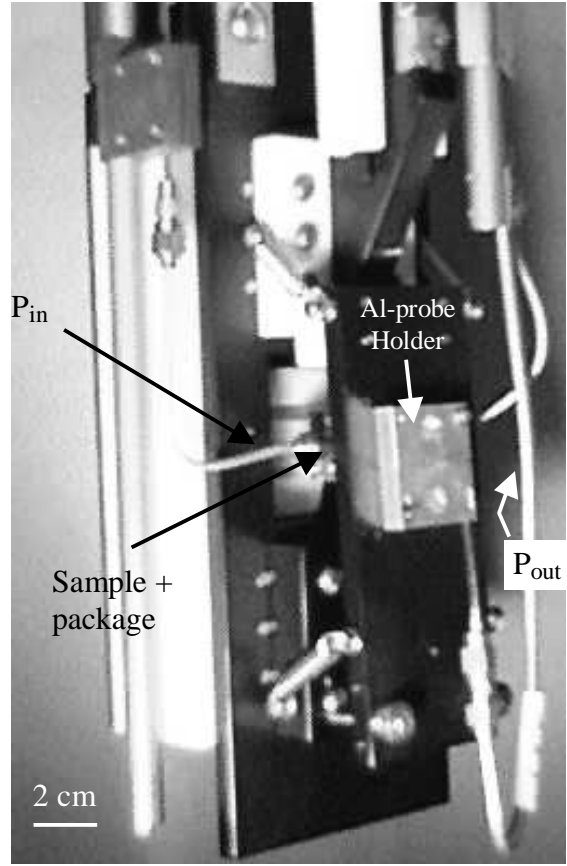

Fig. 2. Photograph of the cryogenic part of the microwave microscope.

A photograph of the package and the sample is shown in Fig. 3. The capacitive coupling pins can be seen on both sides. For imaging, we removed the pin shown on the right hand side, allowing us to image beyond the end of the microstrip. With this arrangement, the resonator is open-ended on both sides, implying that the voltage, and therefore the electric field, will be maximal at the ends of the microstrip for all resonant modes. Before imaging the microstrip, we measured $S_{21}$ using a probe with a 200 µm diameter inner conductor.

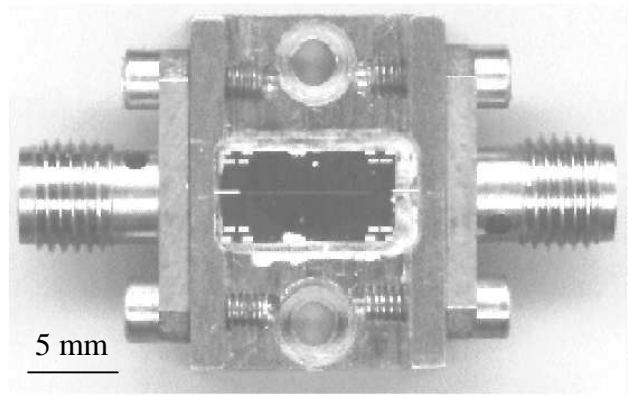

Fig. 3. Photograph of the Cu package with the TBCCO resonator.

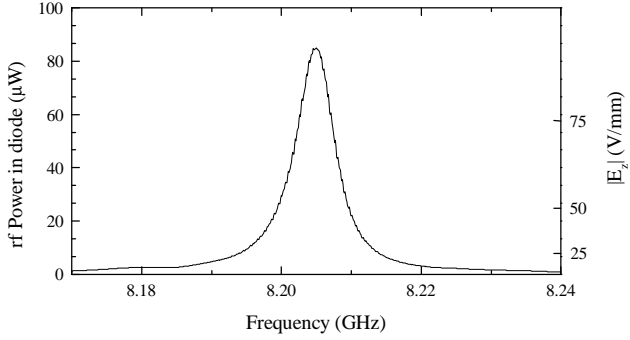

Fig. 4. Frequency response of the TBCCO microstrip, measured with the probe.

The probe was situated at about 250 μm above the open end of the microstrip opposite to the drive pin. The resulting frequency sweep, measured at 77 K, is plotted in Fig. 4. The resonant frequency is approximately 8.205 GHz, while the loaded quality factor is about 1200.

The absolute value of the electric field $|E_z|$ (right axis) has been obtained by converting the power measured at the calibrated diode detector using $E_z = [P/\omega^2\epsilon_0^2 A^2 Z_0]^{1/2}$ [7]. Here $Z_0 = 50\ \Omega$ is the characteristic impedance of the transmission line and $A \approx 0.03$ mm$^2$ is the area of the inner conductor of the probe.

## IV. IMAGES TAKEN AT DIFFERENT HEIGHTS

In order to understand the electric fields generated by the microstrip, we imaged the TBCCO resonator at several different heights. In addition, such an investigation will provide useful information about the perturbation of the resonator by the probe.

Figs. 5(a)-(c) show images taken heights of 1000, 500 and 180 μm, all obtained at a fixed frequency of 8.205 GHz. As expected, the magnitude of the electric field increases rapidly (from about 2.6 V/mm to 90 V/mm) as the probe-sample distance decreases. From the scatter in the electric field image shown in Fig. 5(a), we estimate that the noise in the system is about 0.3 V/mm at 10 GHz for an averaging time of 30ms/pixel.

Besides the quantitative changes with height, there are also several qualitative features worth noticing in Fig. 5. First, none of the images shows the simple $\cos^2(x\pi/L)$ dependence one might naively expect for a microstrip. This is mainly due to the very large aspect ratio of the microstrip and the fact that we are sensitive to the electric field, rather than voltage. The electric field tends to be concentrated at the ends of the narrow microstrip; this is clearly seen in the image taken at the closest separation. For the other images, the height is considerably larger than the inner conductor diameter. The height thus determines the spatial resolution and therefore the images appear to be smoother.

Intuitively one might expect that, as long as the height is larger than the inner conductor diameter, the electric field image will sharpen as the probe approaches the device. However, the peaks observed in the 180 μm image appear to be more sharply peaked than the 200 μm spatial resolution expected for our probe. This feature is due to the perturbation of the resonant frequency by the probe, as we will see below.

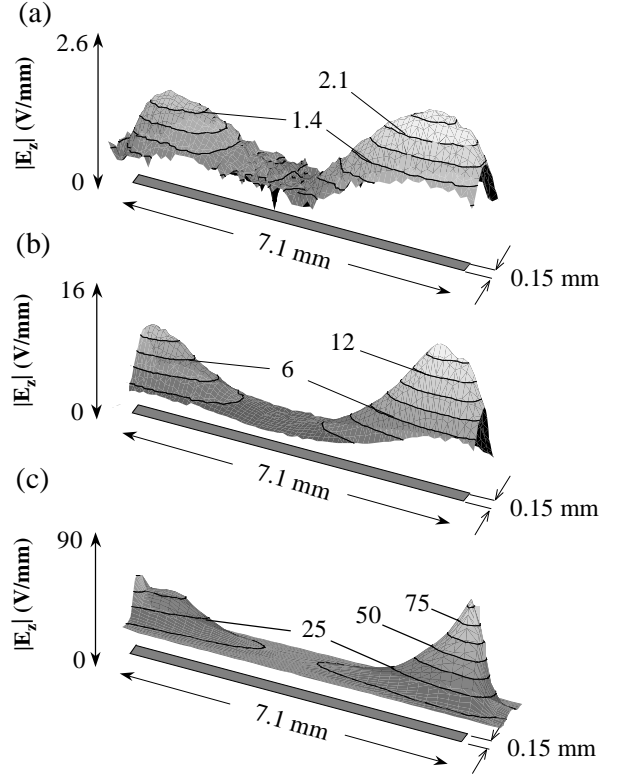

Fig. 5. Absolute value of $E_z$ above an operating Tl$_2$Ba$_2$CaCu$_2$O$_8$ microstrip resonator, measured at fixed frequency at probe heights of (a) 180 μm, (b) 500 μm and (c) 1000 μm. The contour lines are labeled in V/mm, and the microwave power is supplied on the left hand side.

The second noteworthy feature is the curvature of the peaks. While the peaks show a concave curvature at the smallest heights, indicating a very rapid decrease in amplitude away from the antinode, they have a convex curvature in the image taken at a height of 1000 μm. This effect is also due to the perturbation of the high-Q resonance by the probe. This effect is less severe at larger heights, where the probe perturbation is less significant. We would like to stress that these features could be easily overlooked, for example when a 2D grayscale plot is used.

## V. $E_Z$ IN FREQUENCY-FOLLOWING MODE

To determine the extent to which perturbation of the resonant frequency affects our images, we imaged the absolute magnitude of the electric field from the microstrip in the frequency-following mode. We focused on the antinode on the right hand side of the resonator, since it was not possible to image the left antinode entirely due to the presence of the pin.

Fig. 6(a) shows the result, measured at a height of 230 μm, while Fig. 6(b) shows the corresponding image measured at a fixed frequency. In comparison to the fixed-frequency image, $|E_z|$ shows a different curvature, producing a less peaked antinode in the frequency following mode. The difference in cur-vature is very similar to the difference observed between the images taken at small (180 µm) and large (1000 µm) separations in the fixed frequency mode. Both observations can be explained by realizing that the probe creates an electric field perturbation at the ends of the microstrip, which tends to lower the resonant frequency of the device [9]. Since Fig. 6(b) was taken at a fixed frequency that was determined at the center of the antinode, at that particular position the system is on resonance. When the probe is moved to a different position, where the field is lower, the perturbation is reduced, shifting the resonant frequency up. As a result, the measured $|E_z|$ drops considerably faster than the actual field values. This effect will be much more evident at closer separations, where the perturbation is largest.

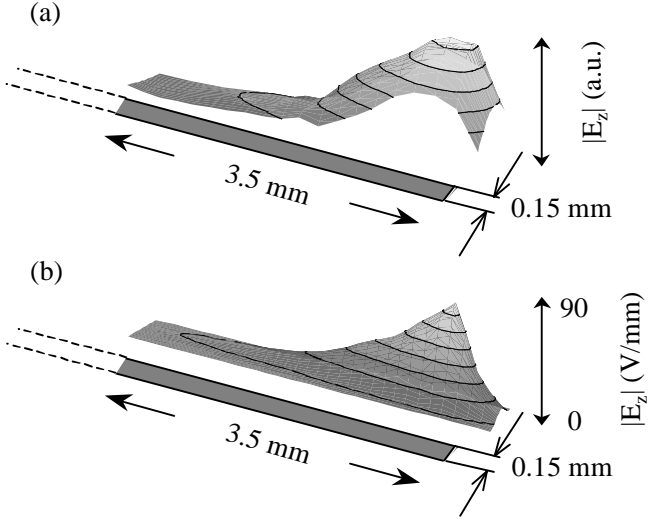

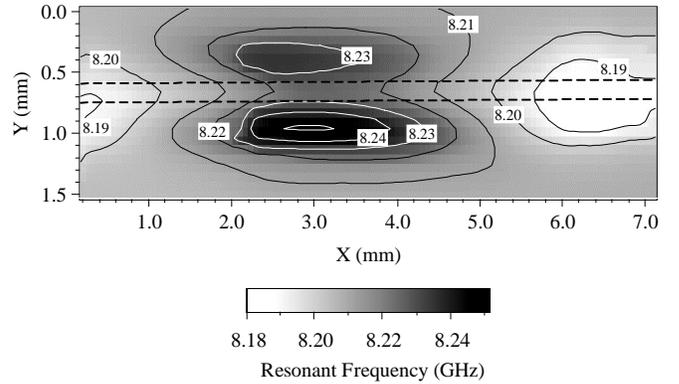

Fig. 7. Resonant frequency measured as a function of probe position at 180 µm above the TBCCO resonator. The dashed lines indicated the edges of the microstrip and the contour lines are labeled in GHz.

Fig. 6. Comparison of the electric field amplitude measured (a) in the frequency-following mode and (b) at a fixed frequency, 8.205 GHz. We measured (a) at h = 230 µm and (b) at 180 µm.

## VI. RESONANT FREQUENCY MAPPING

We also monitored the resonant frequency of the device while scanning the probe across the sample. We can use this information to verify whether the differences in Fig. 6(a) and (b) are mainly due to the shifting of the high-Q resonance. Fig. 7 shows a plot of the absolute frequency vs. probe position. As expected, the resonant frequency shifts considerably, being smallest at the electric field antinodes (~ 8.18 GHz). We also note that the largest frequencies occur when the probe is over the center of the strip. In fact, we can compare the frequencies in near the center (8.24 GHz) with the value measured without the probe present, 8.225 GHz. This means that in the middle of the strip the resonant frequency is, in fact, higher than the unperturbed value. This is presumably due to the fact that at this position, there is an antinode in the magnetic field. Therefore, the dominant effect at this position will be a magnetic field perturbation, which should increase the resonant frequency [9].

## VII. CONCLUSIONS

We have developed a cryogenic near-field microwave microscope that images the local electric field distribution around operating superconducting resonant devices. At small heights the spatial resolution is about 200 µm; this can be improved significantly by using a smaller probe, without reaching the sensitivity limit of the system. We have employed a frequency-following technique to determine the perturbation on the resonant properties of the device caused by the probe. We see clear evidence for perturbations and find that, at least to some extent, we can correct our images for these contributions.

## REFERENCES

[1] Z-Y. Shen, "High-Temperature Superconducting Microwave circuits," Artech House, Boston, 1994.
[2] D. E. Oates, P. P. Nguyen, G. Dresselhaus, M. S. Dresselhaus, C. W. Lam, and S. M. Ali, "Measurement and modeling of linear and nonlinear effects in striplines," J. Supercond. **5**, pp. 361-369, 1992.
[3] G. Hampel, B. Batlogg, K. Krishana, N. P. Ong, W. Prusseit, H. Kinder, and A. C. Anderson, "Third-order non-linear microwave response of $YBa_2Cu_3O_{7-\delta}$ thin films and single crystals," Appl. Phys. Lett. **71**, pp. 3904-3906, December 1997.
[4] T. P. Budka, S. D. Waclawik, and G. M. Rebeiz, "A coaxial 0.5-18 GHz near electric field measurement system for planar microwave circuits using intergrated probes," IEEE Trans. Microwave Theory Tech. **44**, pp. 2174-2184, December 1996, and references therein.
[5] D. W. van der Weide and P. Neuzil, "The nanoscilloscope: combined topography and ac field probing with a micromachined tip," J. Vac. Sci. Technol. B **14**, pp. 4144-4147, November/December 1996.
[6] A. S. Hou, F. Ho, and D. M. Bloom, "Picosecond electrical sampling using a scanning force microscope," Electron. Lett. **28**, pp. 2302-2303, December 1992.
[7] Ashfaq S. Thanawalla, S. K. Dutta, C. P. Vlahacos, D. E. Steinhauer, B. J. Feenstra, Steven M. Anlage, F. C. Wellstood, and Robert B. Hammond, "Microwave near-field imaging of electric fields in a superconducting microstrip resonator," in press.
[8] D. E. Steinhauer, C. P. Vlahacos, S. K. Dutta, F. C. Wellstood, and Steven M. Anlage, "Quantitative imaging of sheet resistance with a scanning near-field microwave microscope," Appl. Phys. Lett. **71**, pp. 1736-1738, February 1997.
[9] L. C. Maier, and J. C. Slater "Field strength measurements in resonant cavities," J. Appl. Phys. **23**, 68 (1952).